\documentclass[aps,prl,reprint,showpacs,floatfix,superscriptaddress]{revtex4-1}
\usepackage{import}
\usepackage{graphicx}
\usepackage{amsmath,amsthm,amsfonts,amssymb,amscd}
\usepackage{hyperref}
\usepackage{xcolor}
\usepackage{xwatermark}
\usepackage{psfrag}
\hypersetup{
     colorlinks   = true,
     citecolor    = blue,
     urlcolor     = blue
}
\usepackage[latin1]{inputenc}

\begin{document}
\title{Description of the associated Legendre functions in analogy to the state space\\ of the free electromagnetic field}
\author{J. Bola\~nos-Coral}\email{jbolanosc@if.ufrj.br}
\affiliation{Instituto de F\'isica, Universidade Federal do Rio de Janeiro, IF-UFRJ.\\Av. Athos da Silveira Ramos 149 - Bl. A, CEP: 21941-972, Rio de Janeiro, RJ, Brasil}
\author{C. E. Cede\~no M.}
\affiliation{Centro Brasileiro de Pesquisas F\'isicas, CBPF. Rua Dr. Xavier Sigaud 150 - Urca,\\ CEP: 22290-180, Rio de Janeiro, RJ, Brasil}
\date{\today}

\begin{abstract}
Currently, some approaches to the associated Legendre functions based on different factorization methods are known. However, they have not allowed identifying new properties that permit to improve our knowledge of any physical system. In this letter, we show that the set of all the associated Legendre functions can be understood in analogy to the state space of the free electromagnetic field. Thanks to this correspondence we hope that any system, classical or quantum, described by such set of functions can be physically understood by using this analogy. We illustrate our results showing that the classical multipole expansion of the scalar and vector potentials is connected to the quantum mechanics discreteness property.
\end{abstract}

\maketitle 
\textit{Introduction.---} The quantum unidimensional harmonic oscillator (1DHO) is a system of clear relevance in different areas of physics \cite{Menicucci, Scully, Scully1}. This kind of systems obeys Schr\"odinger's equation with the following Hamiltonian
\begin{equation}\label{Hamiltoniano}
\mathcal{H}_{1DHO} = -\dfrac{\hbar^{2}}{2m}\dfrac{d^{2}}{dx^{2}} + \dfrac{1}{2}m\omega^{2}x^{2},
\end{equation} 
where $\hbar$ is the reduced Planck constant, $m$ is the oscillator mass, $\omega$ is a constant with dimensions of frequency and $x$ is an unidimensional position operator \cite{Schrodinger}.  

The solutions of the Schr\"odinger's equation for the 1DHO can be determined by using, for example, the ladder operator method \cite{Fock0}. This approach is based on the definition of the creation and annihilation operators, $a^{+}$ and $a^{-}$ respectively. Their eigenstates and corresponding eigenenergies can be written as
\begin{equation}\label{Funciones}
\vert n\rangle =
\prod_{j=1}^{n}\dfrac{a^{+}}{\sqrt{j}}\vert 0\rangle,\qquad
E_{n} =  \hbar\omega\left(n + \dfrac{1}{2}\right),
\end{equation}
here, $n$ = 0, 1, 2, $\cdots$, and the ground eigenstate $\vert 0\rangle$ is defined through the annihilation equation $a^{-}\vert 0\rangle$ = $0$ \cite{Comentario}. We note that Eqs. \eqref{Funciones} are also valid for $n$ = 0, in which case the product is reduced to the identity operator \cite{Lang}. 

Physically, $n$  identifies the zeros or node numbers of $\psi_{n}(x)$ = $\langle x\vert n\rangle$. The nodeless wavefunction $\psi_{0}(x)$ is called as the ground wavefunction. In the same way, any of such wavefunctions with zeros is called as an excited wavefunction of the 1DHO. The ground and the four first excited wavefunctions of the 1DHO are illustrated in Fig. \ref{One}. There, it is easy to observe that the number of zeros coincides with the quantum number $n$.

The ladder operator method provides the 1DHO solutions, that were essentials for the current understanding of the free electromagnetic field (FEF) \cite{Dirac, Glauber}. The state space for such system is considered as a set of $\kappa$-modes, which amplitudes behave in analogy to the coordinates of an assembly of 1DHOs \cite{Glauber}. In analogy to Eqs. \eqref{Funciones}, the  eigenstates for each FEF $\kappa$-mode and their corresponding energies can be written as \cite{Comentario2} 
\begin{equation}\label{Funciones2}
\vert n_{\kappa}\rangle_{\kappa}  =
\prod_{j_{\kappa}=1}^{n_{\kappa}}\dfrac{a_{\kappa}^{+}}{\sqrt{j_{\kappa}}}\vert
0\rangle_{\kappa},\qquad
E_{n_{\kappa}} =  \hbar\omega_{\kappa}\left(n_{\kappa} + \dfrac{1}{2}\right),
\end{equation}
where $n_{\kappa}$ = 0, 1, 2, $\cdots$, and the $\kappa$-ground eigenstate $\vert
0\rangle_{\kappa}$ is defined through the annihilation equation $a_{\kappa}^{-}\vert 0\rangle_{\kappa} = 0$ \cite{Glauber}. Then, any FEF eigenstate can be obtained from the FEF vacuum that corresponds to the tensorial product of all annihilable eigenstates $\vert 0\rangle_{\kappa}$.

The fact that the amplitudes of each mode behave analogously to the coordinates of an assembly of 1DHOs, is a direct consequence of that both, electric and magnetic components, satisfy a wave equation. Then, it is natural to ask if there are relationships, at least analogue to Eqs. \eqref{Funciones2}, that could be found as solutions to a differential equation other than Schr\"odinger's equation. This possibility could show us that systems described for such differential equation can be reinterpreted in analogy to the FEF. 

In this letter, we show that the ``state space" conformed by all associated Legendre functions (ALFs), $P_{\ell}^{m}$  \cite{Legendre, Arfken}, can be understood in analogy to the FEF state space. The ALF state space consist of a set of $\ell$-modes which ``mode amplitudes" behave in analogy to the FEF $\kappa$-mode amplitudes. In order to do that we determine the ``annihilable ALFs" and the adequate creation operators that allow to construct any ``excited ALF", with the aim to find a similar relationship to those to the eigenstates in Eqs. \eqref{Funciones2}.

\begin{widetext}
	\onecolumngrid
	\begin{figure}[h!]  
		\includegraphics[width=18cm]{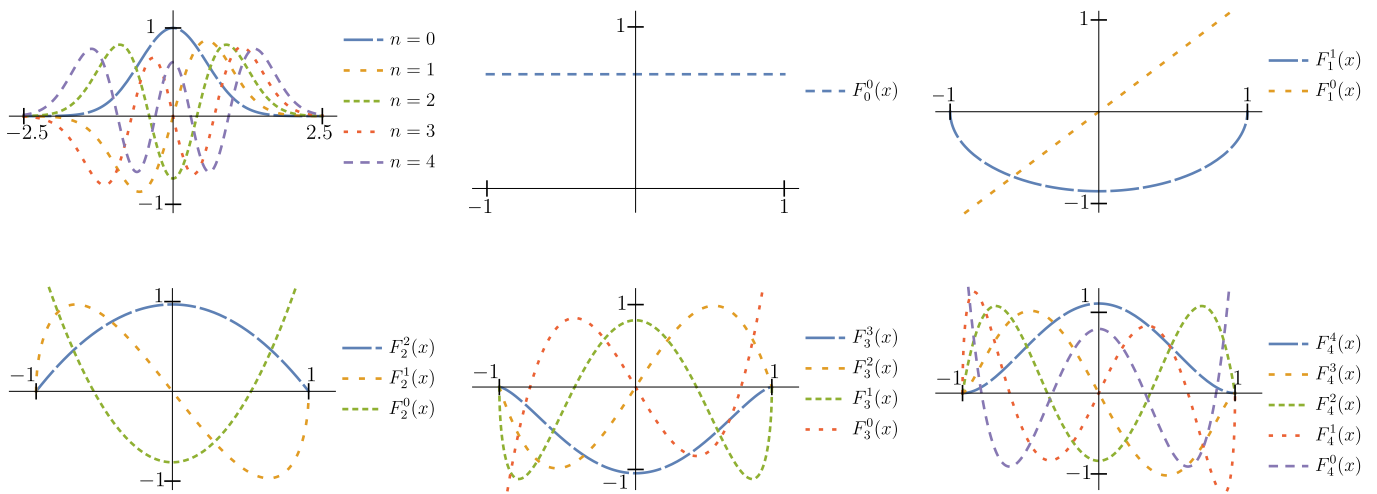} 
		\caption{(Color). The first graph shows the ground and the four first excited wavefunctions of the 1DHO. All these functions are defined in the domain $(-\infty, \infty)$ and have $n$ zeros. The last five plots illustrate the first five ``modes" of the modified ALFs, $F_{\ell}^{m}$. These modes correspond to sets of modified ALFs with the same $\ell$ value. Zeros correspond to intersections with the $X$-axis considering only $\vert x\vert < 1$. In all graphs, there exists one nodeless ALF (blue); for $\ell > 0$, there also exists one one-node ALF (brown); for $\ell > 1$, there also exists one two-nodes ALF (green); for $\ell > 2$ also exists one three-nodes ALF (red), and so on.
		}
		\label{One}%
	\end{figure}
\end{widetext}	

\textit{Results.---} With the aim to facilitate an adequate graphical comparison, we use a set of ``modified ALFs" $F_{\ell}^{m}$=$\sqrt{(2\ell+1)(\ell-m)!/2(\ell+m)!} P_{\ell}^{m}$. These functions conserve the ALF zeros and obey to the orthonormalization relationship $\int_{-1}^{1} F_{\ell}^{m}F_{\ell'}^{m}dx = \delta_{\ell\ell'}$, where $\delta_{\ell\ell'}$ is the Kronecker delta. In contrast to the orthonormalization relationship of the ALFs \cite{Arfken}, this latter reminds those satisfied by wavefunctions in quantum mechanics. 

In Fig. \ref{One} we show the first five sets of these modified ALFs, in which each set corresponds to a specific $\ell$. From this figure we observe that, for each of these sets there is one ALF that, in analogy to the 1DHO ground wavefunction, is nodeless. We also note that for all sets with $\ell >$ 1 there is one ALF with only one zero, in analogy to the first 1DHO excited wavefunction. In the same way, for all sets with $\ell >$ 2 there is one ALF with two zeros, in correspondence to the second 1DHO excited wavefunction, and so on. 

Therefore, from the viewpoint of the zeros number, we note that as $\ell$ increases in a set of ALFs,  it turns more similar to the set of a 1DHO wavefunctions. This could suggest that in reference to the zeros, the set of \textit{all} ALFs behaves in analogy to the FEF state space. This analogy becomes more evident as the parameter $\ell$ increases. With the aim to explore that, we investigate if for each set of ALFs with the same $\ell$ there exists some relationship analogue to Eqs. \eqref{Funciones2}, in which each ``excited ALF" can be constructed from a ``ground ALF".

In view of this discussion, it seems appropriate to introduce a more natural and transparent notation for the ALFs, $\mathcal{P}_{\ell}^{n_{x}}$, where instead of $m$ the node number $n_{x}$ is explicitly shown. 
It is important to note that an ALF with the same indexes in this latter notation and in the traditional way of labelling, $P_{\ell}^{m}$, do not coincide. 

Using the constraint $n_{x}$ = $\ell$ - $m$, the associated Legendre differential equation becomes
\begin{equation}\label{Ecuaa}
 -\dfrac{d}{dx}\left[ \left(1-x^{2}\right)\dfrac{d\mathcal{P}_{\ell}^{n_{x}}}{dx}\right]-\left[\ell(\ell+1) -\dfrac{(\ell - n_{x})^{2}}{1-x^{2}}\right]\mathcal{P}_{\ell}^{n_{x}} = 0,
\end{equation}
where the possible values of $\ell$ and $n_{x}$, for the physical interesting cases, are $\ell = 0, 1, 2, \cdots$, and $n_{x} = 0, \pm 1, \pm 2, \cdots, \pm \ell$. The relationship between ALFs with positive and negative $m$ values \cite{Arfken} can be directly rewrite in terms of $n_{x}$. Then, we will only take into account ALFs with positive $n_{x}$ values here. The extension of our results to negative values of $n_{x}$ will not be considered.

In order to obtain a true analogy between the set of ALFs and the FEF eigenstates, we show that Eq. \eqref{Ecuaa} is solved by using ladder operators, just as in the 1DHO case. In this way, we note that different factorization methods using differential operators, which could be interpreted as ladder operators, has been relevant in the study of different physical systems, e.g., the Infeld and Hull \cite{Infeld, Infeld2}, the Abraham and Moses \cite{Moses} and the Pursey's method \cite{Pursey} and supersymmetric quantum mechanics \cite{Susy}. In particular, the Infeld and Hull's method \cite{Infeld, Infeld2} and supersymmetric quantum mechanics \cite{Das} were applied to the study of the associate Legendre differential equation. However, none of them was used to show that the ALFs set could be constructed in analogy to the FEF state space. 

With this aim, we initially consider the nodeless ALFs, that are given by
\begin{equation}\label{Aniquil}
\mathcal{P}_{\ell}^{0}(x) = \dfrac{\Gamma[2\ell+1]}{2^{\ell}\Gamma[\ell+1]}\left(1-x^{2}\right)^{\ell/2},
\end{equation}
where $\Gamma$ is the Gamma function. If these functions behave in analogy to the ground wavefunction of the 1DHO, then an operator $\mathcal{A}_{\ell 0}^{-}$ satisfying the annihilation equation $\mathcal{A}_{\ell 0}^{-}\mathcal{P}_{\ell}^{0}$ = 0 should be able to be constructed. The form of the operator of Eq. \eqref{Ecuaa} suggests that the annihilation operator could be written as $\mathcal{A}_{\ell 0}^{-}$ = $\sqrt{1-x^{2}}d/dx  + \mathcal{F}_{\ell 0}$, where $\mathcal{F}_{\ell 0}$ is a function to be determined. Using the annihilation equation we find that $\mathcal{F}_{\ell 0}$ = $\ell x/\sqrt{1-x^{2}}$ so that the annihilation operator results as $\mathcal{A}_{\ell 0}^{-} = \sqrt{1-x^{2}}d/dx +\ell 
x/\sqrt{1-x^{2}}$. We also expect that there exists a creation operator that corresponds to $a_{\kappa}^{+}$, such that the one-node function $\mathcal{P}_{\ell}^{1}$ can be obtained by applying this operator on $\mathcal{P}_{\ell}^{0}$. To determine such operator, we initially consider that any one-node ALF can be written as $\mathcal{P}_{\ell}^{1} = \Gamma[2\ell](2^{\ell-1}\Gamma[\ell])^{-1}x\left(1-x^{2}\right)^{(\ell-1)/2}$. Again, the form of the operator of Eq. \eqref{Ecuaa} suggests that $\mathcal{A}_{\ell 1}^{+} = -\sqrt{1-x^{2}}d/dx$ + $\mathcal{F}_{\ell 1}$, where $\mathcal{F}_{\ell 1}$ must be specified. This last can be done considering the creation equation $\mathcal{P}_{\ell}^{1}\sim \mathcal{A}_{\ell 1}^{+}\mathcal{P}_{\ell}^{0}$. Thus $\mathcal{A}_{\ell 1}^{+} = -\sqrt{1-x^{2}}d/dx +\ell  
x/\sqrt{1-x^{2}}$. Just as in the $a_{\kappa}^{+}$ cases, the creation operator determined here does not normalize adequately the functions $\mathcal{P}_{\ell}^{1}$. Then, the inclusion of a normalization constant $\mathcal{C}_{\ell}^{1}$ such that $\mathcal{P}_{\ell}^{1}$ = $(\mathcal{A}_{\ell 1}^{+}/\sqrt{\mathcal{C}_{\ell}^{1}})\mathcal{P}_{\ell}^{0}$ is necessary. 
Therefore, $\mathcal{C}_{\ell}^{1}$ = $(2\ell +1)(2\Gamma[2\ell])^{-1}\int_{-1}^{1} \left[\mathcal{A}_{\ell 1}^{+}\mathcal{P}_{\ell}^{0}\right]^{2}dx$. 

In contrast with the 1DHO and each FEF $\kappa$-mode, where exists only one creation operator with which it is possible to construct all the eigenstates, the creation operators $\mathcal{A}_{\ell 1}^{+}$ are not enough to construct all excited ALFs. Note that these only can be used to determine the functions $\mathcal{P}_{\ell}^{1}$ from $\mathcal{P}_{\ell}^{0}$. There are no other creation possibilities by using such operators.  However, such objective can be reached by the introduction of a set of creation operators, that are given by   
\begin{equation}
\mathcal{A}_{\ell n_{x}}^{+} = -\sqrt{1-x^{2}}\dfrac{d}{dx} +\dfrac{(\ell+1-n_{x})x}{\sqrt{1-x^{2}}},
\end{equation}
in which the operator $\mathcal{A}_{\ell 1}^{+}$ is only the simplest case. In turn, their corresponding normalization constants are
\begin{equation}
\mathcal{C}_{\ell}^{n_{x}}= \dfrac{(2\ell +1)\Gamma[n_{x}+1]}{2\Gamma[2\ell-n_{x}+1]}\int_{-1}^{1} \left[\mathcal{A}_{\ell n_{x}}^{+}\mathcal{P}_{\ell}^{n_{x}-1}\right]^{2}dx.
\end{equation}
These last relationships are valid for $ n_{x}$ = 1, 2, 3 $\cdots$. With this, the general expression for an excited ALF is given by
\begin{equation}\label{FinalF}
\mathcal{P}_{\ell}^{n_{x}} = \prod_{j=1}^{n_{x}}\dfrac{\mathcal{A}_{\ell
j}^{+}}{\sqrt{\mathcal{C}_{\ell j}}}\mathcal{P}_{\ell}^{0}.
\end{equation}
This expression is completely analogous to Eqs. \eqref{Funciones} and \eqref{Funciones2}. Just as in these cases,  the product is reduced to the identity operator for the zero nodes case \cite{Lang}. 

Eq. \eqref{FinalF} is our principal result. It shows that the ``state space" constituted by all the ALFs can be obtained from an {\textit{ALF vacuum} conformed by all annihilable ALFs, in complete analogy to the FEF case. It is worth noticing that our results are general, in the sense that, they unravel a mathematical property of the ALFs. Therefore, we would like to stress that Eq. \eqref{FinalF} could help to explore new characteristics of any physical system, classical or quantum, that could be described through the ALFs. Next, we illustrate the power and usefulness of our results using as an example a classical system.

\textit{Scalar potential for azimuthal symmetry problems.---} In classical electrodynamics it is known that the scalar potential (in short, potential) obeys Laplace's equation. For azimuthal symmetry problems the polar terms of any potential in spherical coordinates usually are written by using Legendre polynomials (LPs) $P_{\ell}(\cos\theta)$ \cite{Jackson}. They coincide with $\mathcal{P}_{\ell}^{\ell}(\cos\theta)$. Hence, using Eq. \eqref{FinalF}, we can write such potential as
\begin{equation}\label{Ej11}
\Phi(r, \theta) =\sum_{\ell = 0}^{\infty}\left[A_{\ell}r^{\ell}+ B_{\ell}r^{-(\ell+1)}\right]\prod_{j=1}^{\ell}\dfrac{\mathcal{A}_{\ell
j}^{+}}{\sqrt{\mathcal{C}_{\ell j}}}\mathcal{P}_{\ell}^{0}(\cos\theta),
\end{equation}
here, $A_{\ell}$ and $B_{\ell}$ are coefficients that must be determined from the boundary conditions. This relationship shows that the polar terms of the potential always can be constructed from a ALF vacuum. In order to illustrate this, we consider the potential established outside of a metallic sphere of radius $R$ carrying a charge $Q$ placed in an otherwise uniform electric field $E_{0}\hat{z}$. Such potential is known \cite{Jackson} and, in correspondence to  Eq. \eqref{Ej11}, we can write this as
\begin{equation}\label{P1}
\Phi(r, \theta) = \dfrac{k_{c}Q}{r}-E_{0}\left(r - \dfrac{R^{3}}{r^{2}}\right)\dfrac{\mathcal{A}_{\ell
11}^{+}}{\sqrt{\mathcal{C}_{11}}}\mathcal{P}_{1}^{0}(\cos\theta),
\end{equation}
with $k_{c}$ the Coulomb constant. The expression for this potential is constituted by one term associate to a point charge $Q$, another term due only to the external field and a third term corresponding to the induced charge on the sphere. Those terms that correspond to the point and induced charges depend on the ground ALF $\mathcal{P}_{0}^{0}(\cos\theta) = 1$ and the excited ALF $\mathcal{P}_{1}^{1}(\cos\theta)$, respectively. However, as showed in Eq. \eqref{P1}, this latter can also be obtained from a ground ALF, i.e., the two potential terms due to the charges depend essentially on the ALF vacuum. This dependence is general for any kind of azimuthal symmetry problem.

\textit{Multipole expansion for electromagnetic potentials with azimuthal symmetry.---} As mentioned above, the potential given in Eq. \eqref{P1} is constituted by one term due to the external field and two terms that correspond to charge distributions. These last are identified as the monopolar and dipolar components of the multipole expansion for the potential. This suggests that the general multipole expansion of the potential could uncover a general relationship between the potential and the ALF vacuum here introduced. In fact, we will proceed to show that this is the case.

In Fig. \ref{Po1} (left) we illustrate an arbitrary localized charge distribution $\rho(\textbf{r}')$. In this, we define a volume differential $dV'$, localized in a position $\textbf{r}'$, inside the distribution. The potential in the position $\textbf{r}$, external to the distribution, is known \cite{Jackson}. Using Eq. \eqref{FinalF} we write such potential as 
 \begin{equation}\label{ExP}
\Phi = \dfrac{1}{4\pi\varepsilon_{0}}\sum_{\ell =0}^{\infty}\dfrac{1}{r^{\ell+1}}\int dV'\left(r' \right)^{\ell}\rho(\textbf{r}') \prod_{j=1}^{\ell}\dfrac{\mathcal{A}_{\ell
j}^{+}}{\sqrt{\mathcal{C}_{\ell j}}}\mathcal{P}_{\ell}^{0}(\cos\theta'),
\end{equation}
with $\varepsilon_{0}$ the vacuum permittivity and $\theta'$ the angle between $\textbf{r}$ and $\textbf{r}'$. From this we observe that each term for the multipole expansion corresponds to some element that constitutes the ALF vacuum. 

Interestingly, a similar result is applicable to the vector potential. In Fig. \ref{Po1} (right) we illustrate a current circuit of intensity $I$ such that its vector potential does not depend on the azimuthal coordinate. In this, a differential of current is localized in a position $\textbf{r}'$. Using Eq. \eqref{FinalF} the vector potential can be written as
\begin{equation}\label{ExA}
\textbf{A}= \dfrac{\mu_{0}I}{4\pi}\sum_{\ell =0}^{\infty}\dfrac{1}{r^{\ell+1}}\oint d\textbf{l}'\left(r' \right)^{\ell}\prod_{j=1}^{\ell}\dfrac{\mathcal{A}_{\ell
j}^{+}}{\sqrt{\mathcal{C}_{\ell j}}}\mathcal{P}_{\ell}^{0}(\cos\theta'),
\end{equation}
where $\mu_0$ is the vacuum permeability. 
Similarly to the scalar potential, each term of this multipole expansion corresponds to some element that conforms the ALF vacuum. 
\begin{figure}[tb]
\begin{tabular}{cc}
\includegraphics[width=4.2cm]{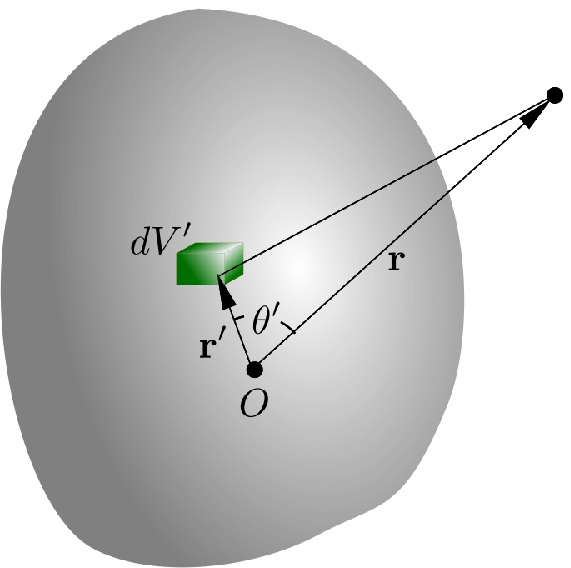} 
&
\includegraphics[width=4.2cm]{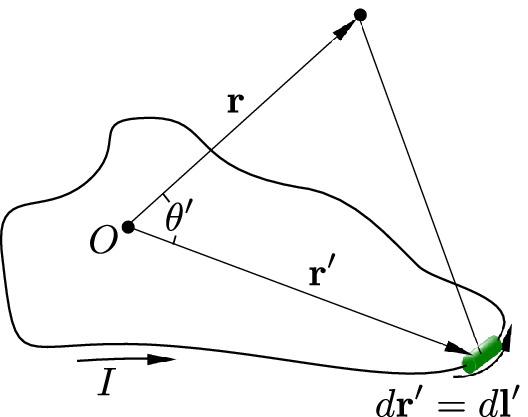}
\end{tabular}

\caption{(Color). Generic charge (left) and current (right) distributions. The corresponding differentials of volume $dV'$ and circuit $d\textbf{l}'$ are localized in a position $\textbf{r}'$ on their frame of reference, respectively. We are interested in the values of the scalar and vector potentials in a point, localized in a position $\textbf{r}$, external to each distribution.}
\label{Po1}%
\end{figure}
In order to visualize the scope of Eq. \eqref{FinalF}, we note that this and Eqs. \eqref{Funciones2} show two indexes that can be compared. In the first appear the indexes $\ell$ and $n_{x}$; whereas that in the second relationship appear $\kappa$ and $n_{\kappa}$ that identify the $\kappa$-modes and the eigenstates, respectively. Although the significance of $n_{x}$ as the number of zeros was discussed, the meaning of $\ell$, at least in this context, is not so clear. However, the analogy here introduced will allow us to identify the precise sense of that.

To simplify our presentation, we will consider the index $\kappa$ in Eqs. \eqref{Funciones2} as associated only to the wave number. In that equation $\kappa$ can take any positive value. However, for the non-free electromagnetic fields case, e.g.,  inside of a cavity, such index can take only some possible values \cite{Cavidades,Scully}. Physically this implies that only photons with determined momentum values can exist. Thus, we say that the mathematical discreteness of $\kappa$ indicates some kind of physical discreteness, where this latter concept must be understood as in quantum mechanics \cite{Ballentine}. On the other hand, in the context of the present work, the analogy here discussed suggests that the discreteness of $\ell$, related with $\kappa$, also should indicates some kind of physical discreteness, even in the classical systems description. 

\textit{Conclusions.---}
We showed that the set of all ALFs can be constructed by using ladder operators similarly to the FEF state space. From that, we conclude that any system described by using such functions, could be understood in analogy to the FEF. Thus, our results allow a reinterpretation of some properties of classical systems in a form closely related to the  quantum mechanics discreteness property. From this viewpoint, our results also could pave the way to improve the understanding of the classical-quantum transition \cite{Zurek}. Finally, given that ALFs are also used to describe quantum systems, we think that our results will allow to identify new characteristics of such kind of systems.

\textit{Acknowledgements.---} This research was fully supported by
the Government of Brazil through the CAPES and CNPq funding agencies.

\end{document}